\documentclass[conference]{IEEEtran}
\usepackage{acronym}  
\usepackage{algpseudocode}
\usepackage{graphicx}
\usepackage{amsmath}
\usepackage{amssymb }
\usepackage{graphicx}
\usepackage[caption=false,font=small,labelfont=sf,textfont=sf]{subfig}
\usepackage{amsmath}
\usepackage{amssymb}
\usepackage{amsthm}
\usepackage{lipsum}
\usepackage{xcolor}
\usepackage[linesnumbered,ruled,vlined]{algorithm2e}
\usepackage[2]{editing}

\theoremstyle{definition}

\usepackage[ruled,vlined]{algorithm2e}
\def\BibTeX{{\rm B\kern-.05em{\sc i\kern-.025em b}\kern-.08em
    T\kern-.1667em\lower.7ex\hbox{E}\kern-.125emX}}

\usepackage[shortcuts,acronym]{glossaries}

\makeglossaries 

\usepackage{geometry}
\def\BibTeX{{\rm B\kern-.05em{\sc i\kern-.025em b}\kern-.08em
    T\kern-.1667em\lower.7ex\hbox{E}\kern-.125emX}}

\begin{document}

\title{Exploring O-RAN Compression Techniques in
Decentralized Distributed MIMO Systems: Reducing Fronthaul
Load}
\author{ Mostafa~Rahmani\textsuperscript{1},
        Junbo Zhao\textsuperscript{1}, Vida Ranjbar\textsuperscript{2}, Ahmed Al-Tahmeesschi\textsuperscript{1}, \\Hamed Ahmadi\textsuperscript{1}, Sofie Pollin \textsuperscript{2}, Alister~G.~Burr\textsuperscript{1}
        \\
    \IEEEauthorblockA{\textsuperscript{1}School of Physics, Engineering and Technology, University of York,\\ \textsuperscript{2}Department of Electrical Engineering, KU Leuven\\
     \\
    }

\thanks{The work presented in this paper was funded by the UK Department for Science, Innovation and Technology under project YO-RAN.}}

\markboth{\today}%
{Shell \MakeLowercase{\textit{et al.}}: Bare Demo of IEEEtran.cls for IEEE Journals}

\maketitle
\begingroup\renewcommand\thefootnote{}
\footnotetext{The work presented in this paper was funded by the UK Department for Science, Innovation and Technology under projects YO-RAN and DU-Volution.}
\endgroup

\begin{abstract}
This paper explores the application of uplink fronthaul compression techniques within Open RAN (O-RAN) to mitigate fronthaul load in decentralized distributed MIMO (DD-MIMO) systems. With the ever-increasing demand for high data rates and system scalability, the fronthaul load becomes a critical bottleneck. Our method uses O-RAN compression techniques to efficiently compress the fronthaul signals. The goal is to greatly lower the fronthaul load while having little effect on the overall system performance, as shown by Block Error Rate (BLER) curves. Through rigorous link-level simulations, we compare our quantization strategies against a benchmark scenario with no quantization, providing insights into the trade-offs between fronthaul data rate reduction and link performance integrity. The results demonstrate that our proposed quantization techniques not only lower the fronthaul load but also maintain a competitive link quality, making them a viable solution for enhancing the efficiency of next-generation wireless networks. This study underscores the potential of quantization in O-RAN contexts to achieve optimal balance between system capacity and performance, paving the way for more scalable and robust DD-MIMO deployments.

\end{abstract}


\section{Introduction}
Distributed Massive MIMO (D-MIMO) is essential for enabling uniform coverage across the next generation wireless networks, utilizing a vast network of distributed antennas or access points (APs) that serve users jointly, thus forming a virtual massive array. This setup ensures macro-diversity, as users are typically near at least one AP, and effectively mitigates cell-edge interference through joint transmission and reception, maintaining consistent signal quality across what were traditionally cell boundaries \cite{haliloglu2023distributed}.

Building on D-MIMO, Cell-Free Massive MIMO (CF-mMIMO) further revolutionizes this concept by completely removing cell boundaries, enabling all APs to coherently serve all users under centralized coordination \cite{rahmani2022deep}. This creates a single cohesive ``cell" throughout the network, eliminating inter-cell interference and turning potential interfering signals from neighboring cells into useful contributions through joint processing. In particular, CF-mMIMO significantly enhances per-user spectral efficiency compared to traditional small cell networks \cite{ngo2017cell} and achieves higher energy efficiency than its co-located massive MIMO counterparts \cite{ahmadi2025towards, mohammadzadeh2025pilot}, depending on the scenario. Despite its advantages, CF-mMIMO encounters significant scalability challenges in a centralized setup. The requirement for all APs to exchange channel state information (CSI) and user payloads with a central processing unit (CPU) leads to unsustainable levels of fronthaul traffic and computational demands. Implementing dedicated high-capacity fronthaul links, such as fiber, becomes prohibitively expensive as the network scales. Moreover, centralized processing can introduce latency, pose reliability risks, and create bottlenecks at a single point of failure \cite{prado2023study}.

To overcome these limitations, more scalable architectures have been developed for CF-mMIMO. User-centric CF-mMIMO, for example, assigns to each user a subset of APs, those with the strongest signal channels, thus reducing coordination overhead. This configuration means APs serve fewer users, and each user is supported by fewer APs, which enhances scalability and eliminates the adverse effects typically experienced at cell edges \cite{bjornson2020scalable}. However, this model often assumes an idealized central control that overlooks practical deployment constraints such as AP hierarchy and physical location.


A more practical and scalable approach is Decentralized Distributed-mMIMO (DD-mMIMO), which organizes the network into clusters of APs managed by local edge processing units (EPUs) \cite{burr2018ultra}. These clusters function as mini CF-MIMO systems, enabling local joint transmission and reception while only requiring limited data exchange between clusters, such as for users near cluster borders. This architecture significantly reduces processing loads and is more aligned with real-world deployments, where APs are grouped with local edge cloud servers. DD-mMIMO fits seamlessly within Open RAN (O-RAN) architectures, where clusters are akin to O-DUs managing groups of O-RUs (APs), with coordination handled by a central O-CU or RAN Intelligent Controller (RIC). The modular and flexible nature of O-RAN, with its emphasis on virtualization and open interfaces, supports distributed cell-free processing effectively \cite{11018310}.

However, even decentralized systems encounter challenges with fronthaul bottlenecks. The transmission of raw or baseband signals across a network can strain the capacity of fronthaul links, particularly when these links utilize wireless or repurposed infrastructure. Employing compression techniques, such as quantizing IQ samples or partially processing data before transmission, can mitigate these loads but also introduces quantization noise, which can degrade signal quality. 

Fronthaul compression, driven by limited link capacity, remains a key challenge in distributed mMIMO networks. Centralized processing of CSI at a CPU enables effective downlink precoding and improved spectral efficiency, but it requires APs to compress CSI for fronthaul transmission. The authors in \cite{silva2023csi} analyzed block compression methods for CSI in distributed mMIMO systems. In \cite{ranjbar2024joint}, cooperative fronthaul quantization was investigated for daisy-chain CF-mMIMO, leveraging correlation between AP signals to reduce quantization noise; however, their analysis considered uncoded bits. Additionally, the authors in \cite{park2024scalable} examined downlink quantization in O-RAN-based CF-mMIMO, proposing a complexity-efficient alternative, $\alpha$-parallel multi-variate quantization ($\alpha$-PMQ), which quantizes precoded signals of low-interference RU clusters in parallel and high-interference clusters sequentially. To further improve scalability, they introduced neural-MQ, replacing exhaustive search with gradient-based optimization. 
 
Additionally, we have contributed to this area through two key studies: in \cite{zhao2025can}, we assessed the validity of Gaussian assumptions for few-bit CSI acquisition in DD-mMIMO, showing a 7.8\% MSE improvement in channel estimation by deriving the true distribution and optimizing quantization parameters. In \cite{mendez2024bler}, we proposed a single-link end-to-end system, showing that optimal quantization yields the block error rate (BLER) performance close to the unquantized case, confirming its practicality in fronthaul-constrained settings. Building upon our previous work \cite{zhao2025can, mendez2024bler}, this paper addresses the critical challenge of fronthaul capacity constraints in DD-mMIMO systems, focusing specifically on the O-RAN 7.2 architecture. Although prior studies have primarily examined the effects of quantization on CSI accuracy and BLER in simplified contexts, the broader implications of fronthaul compression remain largely unexplored.

 We bridge this gap by evaluating how O-RAN-compliant compression techniques affect overall network performance. Using detailed link-level simulations, we analyze trade-offs across modulation schemes and compression settings, specifically comparing three O-RAN-recommended methods. Results show that, even under aggressive compression, performance can approach near-ideal levels—demonstrating viable, scalable paths for cell-free MIMO in next-gen wireless networks.

\section{System Model}
We consider a DD-MIMO system as shown in Fig. 1, where $M_\text{coor}$ RUs, connected to the DU via fronthaul links, serve $K_\text{coor}$ users within the coordination region. However, the DU, which radiates its own coordination region, processes signals only from $K_\text{serv}$ users within the hexagonal service region \cite{zhao2023decentralised}.
We assume that each RU is equipped with $N_r$ antennas, and each user has a single antenna.
\begin{figure}
    \centering
    \includegraphics[width=0.95\linewidth]{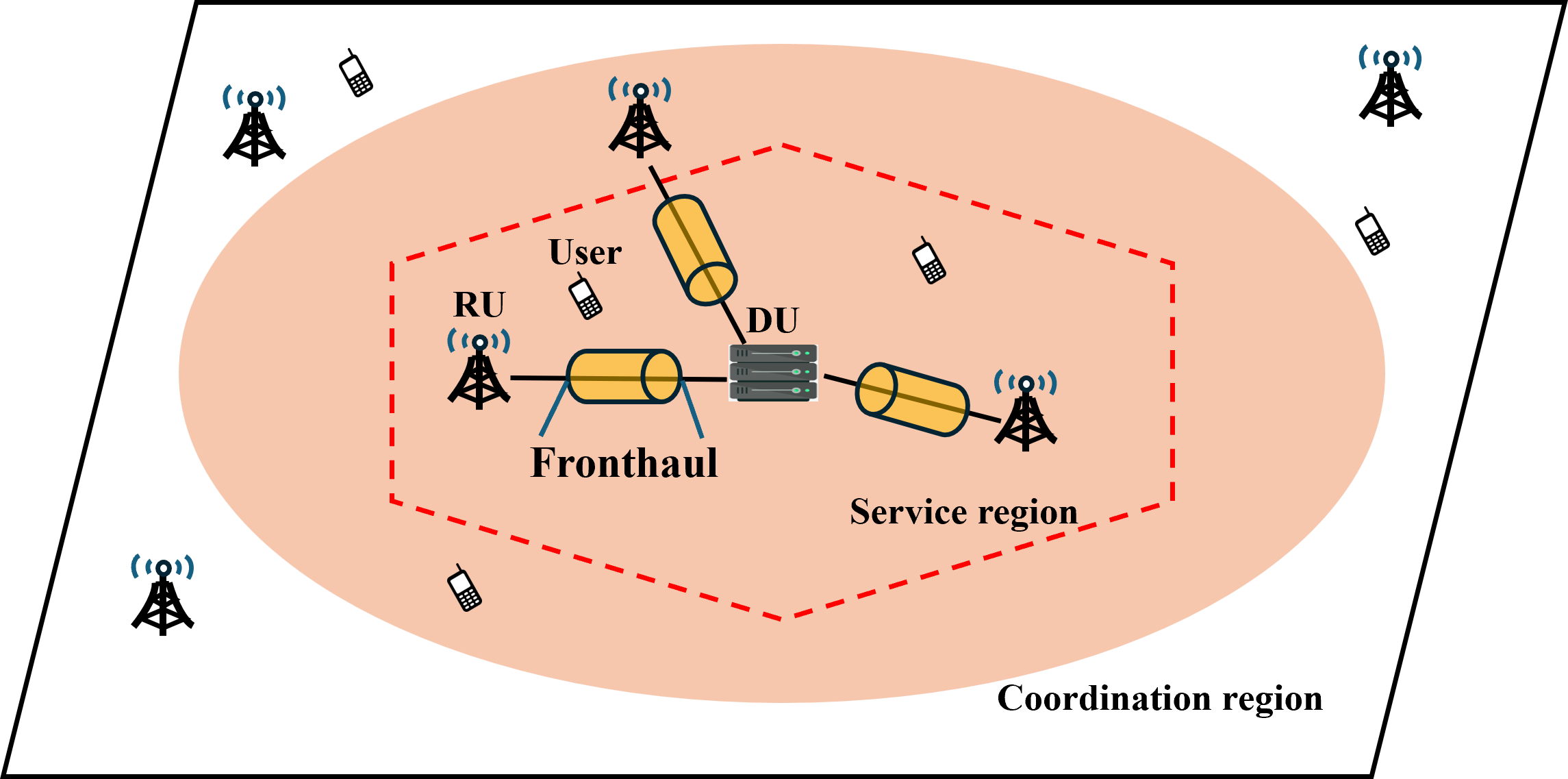}
    \caption{The architecture of DD-mMIMO}
    \label{fig:DD-mMIMO}
    \vspace{-0.5cm}
\end{figure}

To provide a practical analysis, we consider a 5G New Radio (5G-NR) Physical Uplink Shared Channel (PUSCH) chain, as illustrated in Figure 2. The user processes the Uplink Shared Channel (UL-SCH) using low-density parity check (LDPC) coding, followed by PUSCH modulation and OFDM modulation. The data is then transmitted over a multipath channel to RU , where synchronization, OFDM demodulation, and compression are performed. Finally, the compressed data signals from all RUs within the coordination region are transferred to the DU via fronthaul links, where channel estimation, equalization, PUSCH decoding, and LDPC decoding are completed.

We assume that the fading channel is frequency-selective and OFDM is applied in this system. The channel coefficient between the $m$-th RU to the $k$-th user for each resource element (RE) on  the $n$-th subcarrier and $l$-th OFDM symbol $\textbf{g}_{mk}^{(n,l)}\in\mathbb{C}^{N_r}$ is defined as:

\begin{figure*}
    \centering
    \includegraphics[width=1\linewidth]{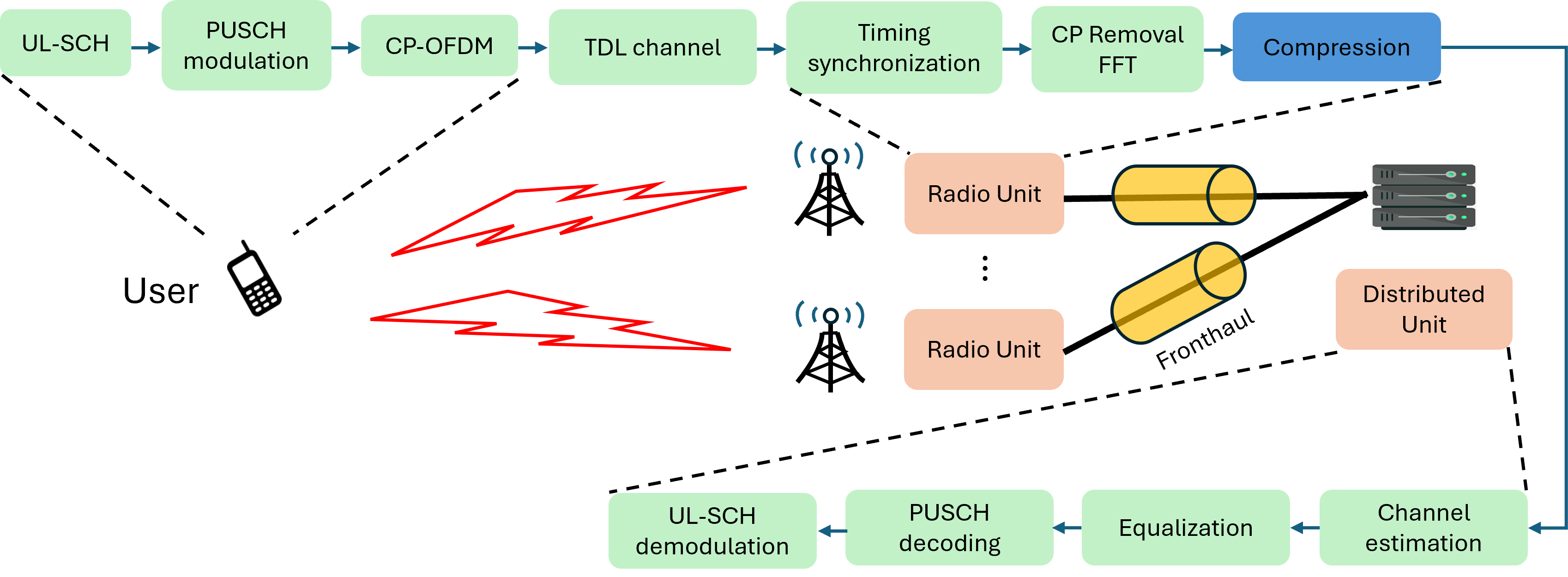}
    \caption{NR PUSCH chain }
    \label{fig:PUSCH}
    \vspace{-0.2cm}
\end{figure*}

\begin{equation}\label{eq1}
    \textbf{g}_{mk}^{(n,l)} = \beta_{mk}^{1/2}\textbf{h}_{mk}^{(n,l)},
\end{equation}
where $\beta_{mk}$ is the large-scale fading coefficient, which only depends on the path loss, $\textbf{h}_{mk}^{(n,l)}$ denotes the small-scale fading in the frequency domain on the $n$-th subcarrier and $l$-th OFDM symbol. In the time domain, we assume the channel model follows a tapped delay line (TDL) with multipath propagation.

The received signal for the RE on the $n$-th subcarrier and the $l$-th OFDM symbol at the $m$-th RU $\textbf{y}_{m}^{(n,l)}\in\mathbb{C}^{N_r}$ is computed by:
\begin{equation}\label{eq2}
    \textbf{y}_{m}^{(n,l)} = \sum_{k=1}^{K_\text{coor}}\textbf{g}_{mk}^{(n,l)}x^{(n,l)}_k + \sum_{k=1}^{K_\text{int}}\textbf{g}_{\text{I},mk}^{(n,l)}x_{\text{I},k}^{(n,l)} + \textbf{z}_m^{(n,l)},
\end{equation}
where $x^{(n,l)}_k$ and $x^{(n,l)}_{\text{I},k}$ denote the data in frequency domain transmitted by the $k$-th user within the coordination region and $k$-th interfering user outside of coordination region, respectively. 
$\textbf{g}_{\text{I},mk}^{(n,l)}$ with subscript `$\text{I}$' represents the channel between the users outside of the coordination region and the RUs within the coordination region. We assume the entities of noise in time domain follow $\mathcal{CN}(0,\sigma_z^2)$. By implementing discrete Fourier transform, the noise at the $m$-th RU in the frequency domain is denoted by $\textbf{z}_m^{(n,l)}$. 

In order to fulfill the limits on fronthaul capacity, the data signal at each RU is quantized before sending it to the DU. The quantized signal (\ref{eq2}) can be expressed in a linear form using Bussgang decomposition \cite{bussgang1952crosscorrelation}:
\begin{equation}\label{eq3}
    \hat{\textbf{y}}_{m}^{(n,l)} = \alpha\textbf{y}_{m}^{(n,l)} + \boldsymbol{\delta}_m,
\end{equation}
where $\alpha$ is a linear coefficient and $\boldsymbol{\delta}_m$ denotes the compression error of the received signal at the $m$-th RU. 

Finally, assuming perfect channel estimation in the DU, we refer to the combining scheme as MMSE. However, since the matrix inversion does not account for quantization noise, this is only an approximate MMSE combiner. This simplification is reasonable given that the channels are already estimated and available at the DU, and the combiner is computed without knowledge of the quantization noise statistics. Under this assumption, the equalized data $\hat{x}_k^{(n,l)}$ for $k = 1,\dots,K_{serv}$, is obtained accordingly.

\begin{equation}\label{eq4}
\begin{split}
    &\hat{x}_k^{(n,l)}\!=\!\textbf{w}^{(n,l)}\hat{\textbf{y}}_{m}^{(n,l)}=\\&\!\!\Bigg(\!\!\sum_{k=1}^{K_\text{coor}}\textbf{g}_{mk}^{(n,l)}\textbf{g}_{mk}^{(n,l)H}\!+\!\sum_{k=1}^{K_\text{int}}\textbf{g}_{\text{I},mk}^{(n,l)}\textbf{g}_{\text{I},mk}^{(n,l)H}\!+\!\sigma_z^2\textbf{I}_{Nr}\!\!\Bigg)^{-1} \textbf{g}_{mk}^{(n,l)^H}\!\!\hat{\textbf{y}}_{m}^{(n,l)},
\end{split}
\end{equation}
where $\textbf{w}^{(n,l)}$ denotes the MMSE combining weight, $\textbf{I}_{Nr}\in\mathbb{R}^{N_r\times N_r}$ is the identity matrix, and we assume the variance of the transmit data $\sigma_x^2=1$.

\section{Fronthaul Compression Techniques in 5G New Radio}
\subsection{5G New Radio Frame Structure}
5G NR employs a flexible frame structure organized hierarchically into radio frames, subframes, slots, and symbols. A radio frame is 10 ms in duration and consists of 10 subframes of 1 ms each. Each subframe can contain one or multiple slots, depending on the chosen numerology. 
Each slot consists of a fixed number of OFDM symbols (typically 14 symbols for normal cyclic prefix) which occupy a portion of the 1 ms subframe (or its fraction) in time. 
In the frequency domain, NR defines REs and resource blocks to describe the allocation of subcarriers. A single OFDM symbol on one subcarrier constitutes the smallest unit called a RE. A group of 12 adjacent subcarriers in frequency over which the transmissions are scheduled, forms a physical resource block (PRB). 
\subsection{Uplink Fronthaul Compression in Open RAN}
O-RAN architectures split the baseband processing between an O-DU and O-RU at a mid-Phy interface (often 3GPP option 7.2x). In this split, the O-RU transmits digitized I/Q samples of the uplink signals over the fronthaul to the O-DU for decoding. The high data rates of 5G (wide bandwidth channels, many antennas, high modulation orders, etc.) can overwhelm the fronthaul capacity if raw I/Q samples are sent without reduction \cite{lagen2021modulation}. The need for compression stems from the limited capacity of the open fronthaul, and the O-RAN Alliance specifies various compression methods to reduce the bandwidth requirement. In fact, for low-PHY splits (Option 7 and even classical C-RAN Option 8), it is critical to apply fronthaul compression to adapt the transport data rate to what the network can support. The goal is to reduce the fronthaul load while minimizing loss of information. The O-RAN specifications (WG4 fronthaul CUS plane) define three primary compression algorithms for uplink IQ data: Block Floating Point (BFP), Block Scaling (BS), and $\mu$-law compression \cite{alliance2022control}. All three algorithms are lossy block compression techniques, meaning some quantization error is introduced during compression. Each compression algorithm processes IQ data blocks corresponding to a PRB, where each PRB consists of 12 IQ samples. Additionally, each method involves sending a small compression parameter per block along with the compressed IQ samples. Below, we explain each compression algorithm in detail.
\subsubsection{\textbf{Block Floating Point (BFP) Compression}}
BFP compression represents each block of IQ samples using a shared floating-point scale. Each IQ sample is compressed to a fixed-size mantissa (e.g., 8 bits), with a shared exponent (e.g., 4 bits) that captures the block's dynamic range. The exponent is determined from the largest magnitude value in the PRB as follows:
\begin{equation}
    e = \text{log}_2(\max_{1\leq i\leq 12}|y_i|) , 
\end{equation}
each sample is normalized to obtain its mantissa: $\textrm{ms}_i = y_i/2^e$, then a uniform quantizer with $m$ bits is applied to $\textrm{ms}_i$:
\begin{equation}
    Q(\textrm{ms}_i) = \Delta_m\cdot\text{round}(\frac{\textrm{ms}_i}{\Delta_m}).
\end{equation}
Finally, the output of the quantizer can be expressed as
\begin{equation}\label{eq7}
    \hat{y}_i = Q(\textrm{ms}_i)\cdot2^e.
\end{equation}

In fronthaul compression, this method compresses the IQ data by transmitting the shared exponent for the entire block plus the quantized mantissas for each sample. The reduction in bits (by dropping least significant mantissa bits) reduces the data rate.

\subsubsection{\textbf{Block Scaling (BS) Compression}}
BS compression is similar in principle to BFP, in that one parameter per block captures the scale, but instead of a base-2 exponent it uses a linear scale factor (an integer scale) to normalize the block. The IQ samples in a block are represented by post-scaled values and a shared 8-bit scaling factor. Typically, the algorithm finds a factor such that when each sample is divided by this factor, the maximum output fits in the target bit-width (e.g. 8-bit magnitude). This factor (or its inverse) is sent to the decoder. In O-RAN, the scale is quantized to 8 bits in a Q1.7 fixed-point format (one integer bit and 7 fractional bits). Intuitively, BS attempts to use the full range of the output bits by linearly scaling the entire block’s values to span the 8-bit dynamic range, rather than restricting scaling to powers of two. This can reduce quantization error compared to BFP (since the scale can be any value 1–255, not just $2^n$) at the cost of a slightly more complex operation (a multiply/divide instead of a simple bit shift). The shared scale value is included in the fronthaul message for that block.

Simply put, for a specific PRB, the scaling factor $S$ can be computed as
\begin{equation}
    S = \frac{\max_{1\leq i\leq 12}|y_i|}{\lambda},
\end{equation}
where $\lambda$ is a design parameter chosen so that the normalized values lie close to the limits of the quantizer. 

The IQ samples is then scaled by $\bar{y}_i = y_i/S$. Similarly, the uniform quantizer is applied to the normalized block with $L=2^m$ levels
\begin{equation}
    Q(\bar{y}_i) = \text{round}\Bigg[\bar{y}_i\bigg(\frac{L}{2}-1\bigg)\Bigg],
\end{equation}
and clipping the values to the range $[-L/2,L/2-1]$. Finally, the compressed result is given by:
\begin{equation}\label{eq10}
    \hat{y}_i = Q(\bar{y}_i)\cdot\frac{S}{\frac{L}{2}-1}.
\end{equation}

\subsubsection{ $\boldsymbol{\mu}$ \textbf{Law}}
$\mu$-law compression is a non-linear commanding technique originally from audio telephony (ITU-T G.711 standard) that O-RAN repurposes for fronthaul IQ data. Unlike BFP or BS (which are linear scalings), $\mu$-law companding uses a logarithmic-like compression curve so that small signal values are quantized more finely than large values. This exploits the fact that in many signals (including uplink receivers or voice), low-magnitude values are more frequent or more important to preserve precisely than very large values. The name “$\mu$-law” comes from the parameter $\mu$ which controls the curve’s nonlinearity. O-RAN chooses $\mu = 8$ and implements a piecewise linear approximation of the standard $\mu$-law curve. In the O-RAN uplink context, $\mu$-law compression works in three stages: (1) extract the sign of each sample and work with absolute values; (2) apply a bit-wise left shift (a power-of-two scaling) to “normalize” the block’s maximum value to the full scale; (3) apply a piecewise compression mapping to each magnitude value, dividing the input range into three segments with different linear scales. The result is a compressed value (typically 8-bit magnitude + sign, for a total of 9 bits) that favors resolution for smaller amplitudes. A 4-bit shift value is transmitted per block (instead of an exponent or scale) to indicate how much left shifting was applied in stage (2) The decoder will reverse the steps: undo the piecewise companding (expanding according to the inverse nonlinear law) and then right-shift the values back by the same amount, reapplying the sign to get the reconstructed samples. 
For a normalized received signal y with $|y|\leq1$, the $\mu$-law compression function is defined as:
\begin{equation}
    F(y) = \text{sgn}(y)\cdot\frac{\text{ln}(1+\mu|y|)}{\text{ln}(1+\mu)}.
\end{equation}
After compressing $y$ with $F(y)$, the uniform quantization is applied as $Q(F(y))$. To recover the original domain, the inverse $\mu$-law expansion is used:
\begin{equation}\label{eq12}
    \hat{y}= \text{sgn}\bigg(Q\big(F(y)\big)\bigg)\cdot\frac{(1+\mu)^{\big|Q\big(F(y)\big)\big|}-1}{\mu}.
\end{equation}

\subsection{Uniform Quantization as a Baseline Technique}
In addition to the standardized uplink compression techniques described earlier, a simple and intuitive approach that can serve as a baseline or benchmark is uniform quantization. In this method, the input IQ samples are uniformly quantized to a fixed number of bits, typically denoted as $n$. This technique applies a straightforward mapping where the amplitude range of the signal is divided into $2^n$ equally spaced quantization levels. Each IQ sample is then approximated by the nearest quantization level, effectively reducing the bitwidth while preserving the overall signal structure. Although uniform quantization does not exploit signal statistics or dynamic range as effectively as more advanced methods like Block Floating Point or $\mu$-law compression, it provides a useful reference point for evaluating the trade-off between compression efficiency and signal distortion. The uniform quantization function can be mathematically expressed as:

\begin{equation}
  Q(y_i) = \Delta \cdot \text{round}\left( \frac{y_i}{\Delta} \right)  ,
\end{equation}

where $\Delta$ is the quantization step size, and $\text{round}(\cdot)$ denotes rounding to the nearest integer. The step size $\Delta$ is typically calculated based on the input signal’s dynamic range $[y_{\min}, y_{\max}]$ as $\Delta = \frac{y_{\max} - y_{\min}}{2^n}$.
According to the Bussgang decomposition, this non-linear quantizer can be expressed in linear form, as given by (3). Note that the linear coefficient in (3) can be computed in the uniform quantization scenario and can also be used to represent the compressed signal in (\ref{eq7}), (\ref{eq10}), and (\ref{eq12}). However, there is not a way to calculate the linear coefficient in compression approach. Simply put, the quantizer expression is given as
\begin{align}
    \mathbf{Q}(\mathbf{y}) =\ & \Delta \cdot \textit{min}\Biggl( 
        \max\Biggl( 
            \left( \left\lfloor \frac{\mathbf{y}}{\Delta} \right\rfloor + \frac{1}{2} \right),
            -\frac{2^{m}-1}{2} 
        \Biggr), \notag \\
    & \frac{2^{m}-1}{2} 
    \Biggr),
\end{align}
where $\Delta$ is the quantization interval and $\mathit{m}$ is the number of quantization bits. The quantization range is limited to $\pm \frac{\Delta}{2}(2^{m} - 1)$. This quantization process can generally lead to two forms of distortion, known as granular distortion due to quantization error within the range, and overload distortion due to truncation of the signal outside the quantization range. For this work, $\mathit{m}$ and optimum $\Delta$ are chosen for optimal performance.

\section{Simulation Results}
\subsection{Parameters and Setup}
To assess the impact of fronthaul compression, we developed a 3GPP-compliant link-level simulator for a DD-MIMO system with 8 RUs and 2 UEs. This compact setup balances simulation accuracy and computational feasibility, as the full PHY stack—including coding, modulation, and channel effects—is modeled in detail. Large-scale fading is calculated using the 3GPP path loss model combined with uncorrelated shadow fading, following the approach in \cite{rahmani2024error}.

Due to varying uplink path losses in DD-mMIMO, we use transmit SNR (TX-SNR) instead of standard SNR. TX-SNR is the ratio of per-antenna transmit power per resource element to receiver noise power, assuming uniform transmit power and scaled noise. This compensates for high path loss by setting a high enough TX-SNR to ensure accurate system performance evaluation across scenarios.

Performance is measured via BLER vs. TX-SNR curves, where BLER is the ratio of erroneous transport blocks (TBs) to total transmissions. Each TX-SNR point is averaged over $2 \times 10^5$ TBs, with independent TDL channel realizations applied per TB to ensure statistical robustness \cite{rahmani2025securing}. The TDL channel is configured with a TDL-B delay profile, a 30 ns delay spread, 0 Hz Doppler shift, and a noise temperature of 290 K. Furthermore, the MCSs employed in the simulations, which are selected from \cite{rahmani2024error}, are listed in Table \ref{tab_1}. 
\begin{table}[!t]
	\centering
	\caption{MCS and TBS configuration in this paper}\label{tab_1}
	\vspace{-2mm}
	\label{table:mcs}

    \begin{tabular}{|c|c|c|c|c|}
    \hline
        MCS & Modulation Scheme & Code Rate  &  PRBs & TBS \\ \hline

        [1] & QPSK & 120/1024 & 25 & 848 \\ \hline

         [2] & 16QAM & 434/1024 & 25 & 6016 \\ \hline

       [3] & 64QAM & 616/1024 & 25 & 13064 \\ \hline

        [4] &  256QAM & 682.5/1024 & 25 & 18960 \\ \hline
 
    \end{tabular}
\vspace{-4mm}
\end{table}
\begin{figure*}[t!]
\centering
\subfloat[\footnotesize For QPSK and 64QAM ]{
\includegraphics[height=61mm]
{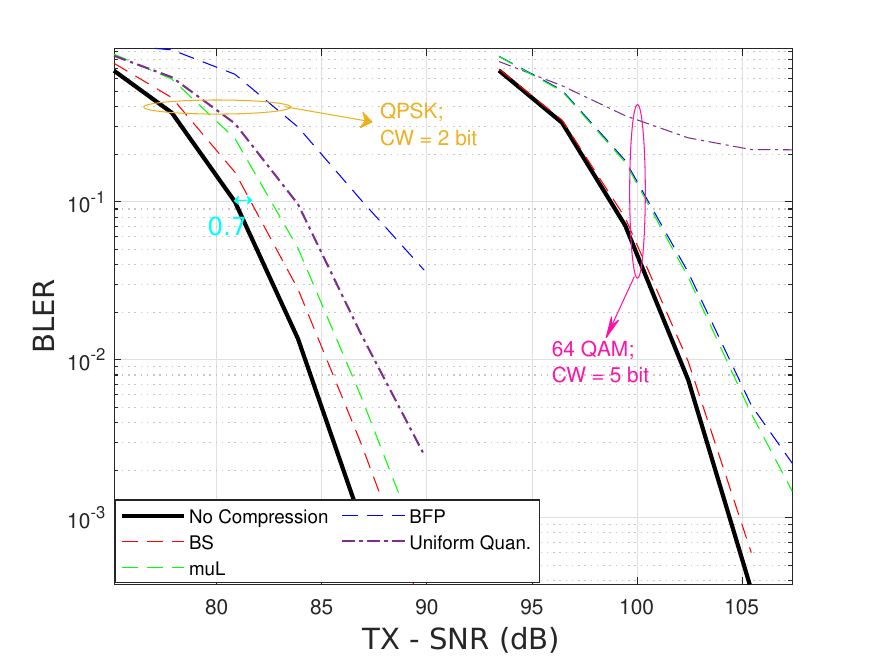}
\label{BLERQPSK}
}
\hspace{10pt} 
\vspace{-.02in}
\subfloat[ \footnotesize For 16QAM and 256QAM ]{
\includegraphics[height=61mm]
{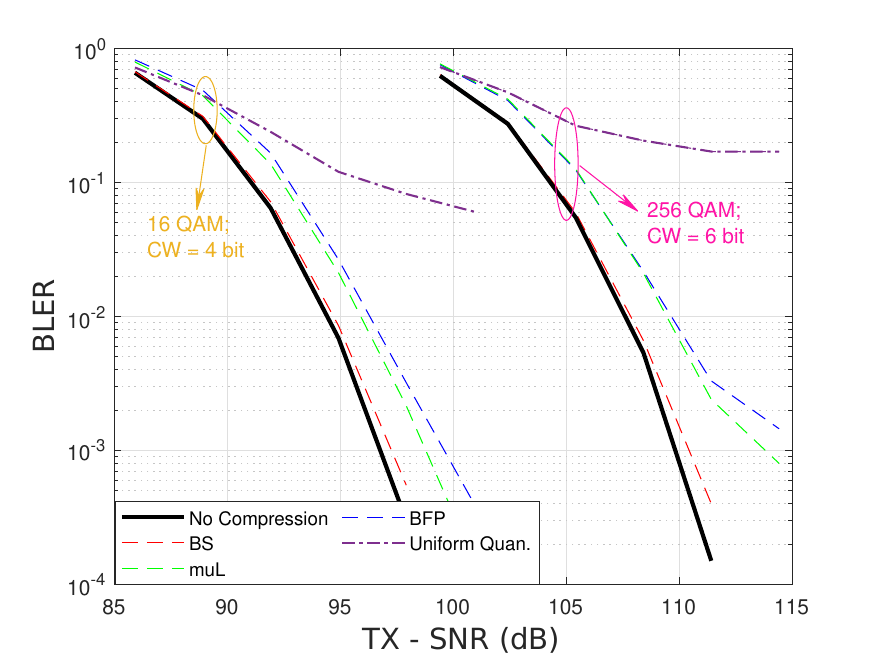}
\label{BLER256}
}
\caption{BLER versus TX-SNR curves for UE1 for DD-MIMO system with 8 RUs and 2 UEs.}

\vspace{-.55cm}
\label{BLER}
\end{figure*}
\subsection{Numerical Results}
We now present numerical results to quantify the effects of fronthaul compression on system performance. 
Figure \ref{BLER} presents the BLER performance versus TX-SNR for UE1, comparing various modulation schemes and compression configurations. The results highlight a clear relationship between modulation order and the required compression bit-width to maintain acceptable performance. Specifically, lower-order modulations, such as QPSK, achieve near-baseline performance even at very low compression bit-widths. For instance, as depicted in Figure \ref{BLERQPSK}, QPSK modulation with 2-bit compression using the BS method achieves performance remarkably close to the uncompressed scenario, requiring 0.7 dB additional SNR to reach a target BLER of $10^{-1}$. This emphasizes the effectiveness of the BS compression method in preserving system performance under aggressive compression settings. For higher-order modulations (64QAM, 16QAM, and 256QAM), as shown in Figure \ref{BLER}, the compression bit-width required to maintain acceptable performance increases, due to greater sensitivity to quantization distortion. Among all evaluated schemes, the BS approach consistently matches the performance of the no-compression benchmark very closely, particularly evident in 64QAM (5 bits) and 256QAM (6 bits) scenarios. Other O-RAN compression methods (BFP and muL) also achieve acceptable performance but show slightly higher SNR differences compared to BS. In contrast, Uniform Quantization clearly demonstrates inferior performance, especially at high modulation orders, exhibiting significantly higher SNR penalties even at relatively large compression widths.

\begin{figure}
    \centering
    \includegraphics[width=1\linewidth]{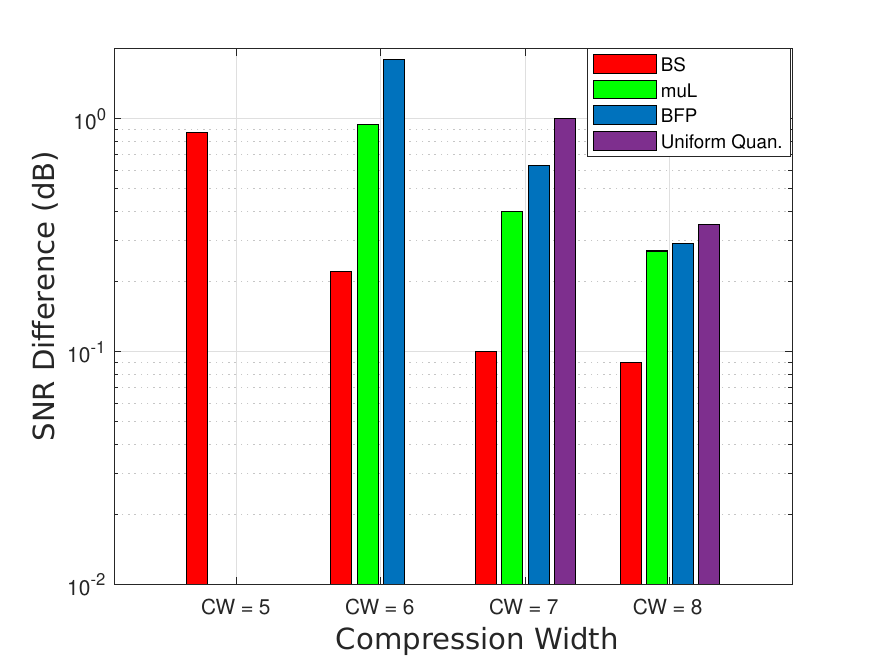}
    \caption{SNR difference for 256 QAM and code rate 682.5 for DD-MIMO system with 8 RUs and 2 UEs. Note: For CW=5, only BS yields acceptable results; for CW=6, Uniform Quantization does not achieve the desired BLER, hence bars are omitted.}
    \label{barchart}
    \vspace{-.5cm}
\end{figure}
Figure \ref{barchart} presents a detailed comparison of the SNR differences for various compression schemes at different compression widths (CW = 5, 6, 7, and 8 bits) for the high-order modulation (256QAM). The SNR difference shown here quantifies how many additional dB are required by each compression method to achieve a BLER target of $10^{-1}$ compared to the no compression benchmark.

As clearly depicted, the BS method consistently achieves the smallest SNR difference across all compression widths, signifying the least performance degradation among all methods. Specifically, at $CW = 5$ bits, BS demonstrates outstanding performance with an acceptable difference, whereas other methods are unable to reach the defined BLER target at this aggressive compression width. For higher compression widths ($CW \geq 6$ bits), although all methods yield acceptable performance, BS consistently provides superior results, followed closely by muL and BFP, while Uniform Quantization shows the largest SNR difference, emphasizing its limited effectiveness in preserving signal quality for high-order modulations. Overall, these results underline the effectiveness and robustness of the BS method in minimizing signal degradation at various compression settings.


\section{Conclusion}
In this work, we have investigated the impact of various fronthaul compression schemes on the performance of a DD-MIMO system with 8 RUs and 2 UEs. Using a detailed, 3GPP-compliant link-level simulator, we compared the effectiveness of three O-RAN compression approaches—BS, muL, and Block BFP—alongside standard Uniform Quantization under multiple modulation schemes and different compression bit-widths (2–8 bits).

Our results clearly demonstrate the necessity of adapting compression settings according to modulation order. Lower-order modulations, such as QPSK, allow aggressive compression (e.g., 2 bits) with minimal performance loss, whereas higher-order modulations, particularly 256QAM, require higher bit-widths (6 bits or more) to avoid significant degradation. Among all tested approaches, the BS method consistently achieves superior performance, exhibiting minimal SNR penalties across all modulation schemes and compression levels. In contrast, Uniform Quantization consistently shows the highest performance degradation, underscoring the importance of structure-aware compression. These findings confirm that appropriate compression algorithms and bit-width selections are critical for efficient and reliable fronthaul design in future distributed MIMO networks.

\IEEEpeerreviewmaketitle

\ifCLASSOPTIONcaptionsoff
  \newpage
\fi


\bibliographystyle{IEEEtran}
\bibliography{refs}

\begin{thebibliography}{10}
\providecommand{\url}[1]{#1}
\csname url@samestyle\endcsname
\providecommand{\newblock}{\relax}
\providecommand{\bibinfo}[2]{#2}
\providecommand{\BIBentrySTDinterwordspacing}{\spaceskip=0pt\relax}
\providecommand{\BIBentryALTinterwordstretchfactor}{4}
\providecommand{\BIBentryALTinterwordspacing}{\spaceskip=\fontdimen2\font plus
\BIBentryALTinterwordstretchfactor\fontdimen3\font minus \fontdimen4\font\relax}
\providecommand{\BIBforeignlanguage}[2]{{%
\expandafter\ifx\csname l@#1\endcsname\relax
\typeout{** WARNING: IEEEtran.bst: No hyphenation pattern has been}%
\typeout{** loaded for the language `#1'. Using the pattern for}%
\typeout{** the default language instead.}%
\else
\language=\csname l@#1\endcsname
\fi
#2}}
\providecommand{\BIBdecl}{\relax}
\BIBdecl

\bibitem{haliloglu2023distributed}
O.~Haliloglu, H.~Yu, C.~Madapatha, H.~Guo, F.~E. Kadan, A.~Wolfgang, R.~Puerta, P.~Frenger, and T.~Svensson, ``Distributed {MIMO} systems for {6G},'' in \emph{2023 Joint European Conference on Networks and Communications \& 6G Summit (EuCNC/6G Summit)}.\hskip 1em plus 0.5em minus 0.4em\relax IEEE, 2023, pp. 156--161.

\bibitem{rahmani2022deep}
M.~Rahmani, M.~Bashar, M.~J. Dehghani, A.~Akbari, P.~Xiao, R.~Tafazolli, and M.~Debbah, ``Deep reinforcement learning-based sum rate fairness trade-off for cell-free {mMIMO},'' \emph{IEEE Transactions on Vehicular Technology}, vol.~72, no.~5, pp. 6039--6055, 2022.

\bibitem{ngo2017cell}
H.~Q. Ngo, A.~Ashikhmin, H.~Yang, E.~G. Larsson, and T.~L. Marzetta, ``Cell-free massive {MIMO} versus small cells,'' \emph{IEEE Transactions on Wireless Communications}, vol.~16, no.~3, pp. 1834--1850, 2017.

\bibitem{ahmadi2025towards}
H.~Ahmadi, M.~Rahmani, S.~B. Chetty, E.~E. Tsiropoulou, H.~Arslan, M.~Debbah, and T.~Quek, ``Towards sustainability in 6g and beyond: Challenges and opportunities of open ran,'' \emph{arXiv preprint arXiv:2503.08353}, 2025.

\bibitem{mohammadzadeh2025pilot}
S.~Mohammadzadeh, M.~Rahmani, K.~Cumanan, A.~Burr, and P.~Xiao, ``Pilot and data power control for uplink cell-free massive {MIMO},'' \emph{arXiv preprint arXiv:2502.19282}, 2025.

\bibitem{prado2023study}
D.~Prado~Alvarez, ``Study of feasible cell-free massive {MIMO} systems in realistic indoor scenarios,'' Ph.D. dissertation, Universitat Polit{\`e}cnica de Val{\`e}ncia, 2023.

\bibitem{bjornson2020scalable}
E.~Bj{\"o}rnson and L.~Sanguinetti, ``Scalable cell-free massive {MIMO} systems,'' \emph{IEEE Transactions on Communications}, vol.~68, no.~7, pp. 4247--4261, 2020.

\bibitem{burr2018ultra}
A.~Burr, M.~Bashar, and D.~Maryopi, ``Ultra-dense radio access networks for smart cities: Cloud-{RAN}, fog-{RAN} and" cell-free" massive {MIMO},'' \emph{arXiv preprint arXiv:1811.11077}, 2018.

\bibitem{11018310}
Y.~Chu, M.~Rahmani, J.~Shackleton, D.~Grace, K.~Cumanan, H.~Ahmadi, and A.~Burr, ``Testbed development: An intelligent {O-RAN}-based cell-free {MIMO} network,'' \emph{IEEE Communications Magazine}, vol.~63, no.~6, pp. 74--81, 2025.

\bibitem{silva2023csi}
M.~Silva, L.~Ramalho, I.~Almeida, E.~Medeiros, and A.~Klautau, ``{CSI} compression for distributed-{MIMO} with centralized precoding and power allocation,'' \emph{IEEE Communications Letters}, vol.~27, no.~6, pp. 1535--1539, 2023.

\bibitem{ranjbar2024joint}
V.~Ranjbar, R.~Beerten, M.~Moonen, and S.~Pollin, ``Joint sequential fronthaul quantization and hardware complexity reduction in uplink cell-free massive mimo networks,'' in \emph{2024 Joint European Conference on Networks and Communications and 6G Summit (EuCNC/6G Summit)}, 2024, pp. 1--6.

\bibitem{park2024scalable}
S.~Park, A.~H. Gokceoglu, L.~Wang, and O.~Simeone, ``Scalable multivariate fronthaul quantization for cell-free massive {MIMO},'' \emph{arXiv preprint arXiv:2409.06715}, 2024.

\bibitem{zhao2025can}
J.~Zhao, M.~Rahmani~Ghourtani, and A.~G. Burr, ``Can we rely on gaussian distribution for few-bit {CSI} acquisition in decentralized distributed massive {MIMO}?'' in \emph{2025 IEEE Wireless Communications and Networking Conference (WCNC)}.\hskip 1em plus 0.5em minus 0.4em\relax IEEE, 2025.

\bibitem{mendez2024bler}
L.~M{\'e}ndez-Monsanto, A.~MacQuarrie, M.~Rahmani~Ghourtani, M.~J.~L. Morales, A.~G. Armada, and A.~Burr, ``{BLER}-{SNR} curves for {5G} {NR} {mcs} under {AWGN} channel with optimum quantization,'' in \emph{2024 IEEE 100th Vehicular Technology Conference (VTC2024-Fall)}.\hskip 1em plus 0.5em minus 0.4em\relax IEEE, 2024, pp. 1--6.

\bibitem{zhao2023decentralised}
J.~Zhao, ``Decentralised distributed massive {MIMO},'' Ph.D. dissertation, University of York, 2023.

\bibitem{bussgang1952crosscorrelation}
J.~J. Bussgang, ``Crosscorrelation functions of amplitude-distorted gaussian signals,'' 1952.

\bibitem{lagen2021modulation}
S.~Lag{\'e}n, L.~Giupponi, A.~Hansson, and X.~Gelabert, ``Modulation compression in next generation ran: Air interface and fronthaul trade-offs,'' \emph{IEEE Communications Magazine}, vol.~59, no.~1, pp. 89--95, 2021.

\bibitem{alliance2022control}
O.~Alliance, ``Control, user and synchronization plane specification,'' \emph{O-RAN Fronthaul Working Group, ORAN-WG4. CUS. 0-v10. 00}, 2022.

\bibitem{rahmani2024error}
M.~R. Ghourtani, J.~Zhao, Y.~Chu, H.~Ahmadi, D.~Grace, R.~G. Maunder, and A.~G. Burr, ``Link-level evaluation of uplink cell-free {MIMO} in {5G NR} over frequency-selective channels,'' \emph{IEEE Open Journal of the Communications Society}, 2024.

\bibitem{rahmani2025securing}
M.~Rahmani~Ghourtani, J.~Zhao, M.~Bashar, K.~Cumanan, A.~G. Burr, and R.~Tafazolli, ``Securing {5G} {NR} networks: Innovative artificial noise methods for protecting cell-free massive {MIMO},'' in \emph{2025 IEEE Wireless Communications and Networking Conference (WCNC)}.\hskip 1em plus 0.5em minus 0.4em\relax IEEE, 2025.

\end{thebibliography}

\end{document}